\documentclass[final, 12pt]{elsarticle}

\usepackage[utf8]{inputenc}
\usepackage[T1]{fontenc}
\usepackage{amsmath,amssymb}
\usepackage{csquotes}  
\usepackage{lineno}    
\usepackage{enumitem}  
\usepackage[nameinlink]{cleveref}
\usepackage{graphicx}  
\usepackage[dvipsnames,table]{xcolor}
\usepackage{url}       

\usepackage{makecell}
\usepackage{fontawesome}
\usepackage{tikz}
\usetikzlibrary{positioning}
\usepackage{forest}

\usepackage{pgfplots}     
\usetikzlibrary{patterns} 
\pgfplotsset{compat=1.15} 

\newcommand{\jeopardy}{jeopardy}
\newcommand{\attackdefense}{attack-defense}
\newcommand{\numtotaltasks}{12,952}
\newcommand{\numtotalwriteups}{23,517}
\newcommand{\numdownloadedwriteups}{15,963}

\journal{Computers \& Security}

\newcommand\copyrighttext{%
  \footnotesize \textcopyright\ 2020. This manuscript version is made available under the CC-BY-NC-ND 4.0 license \url{https://creativecommons.org/licenses/by-nc-nd/4.0/}
  
  Please cite this article as follows: V. Švábenský, P. Čeleda, J. Vykopal, and S. Brišáková, \textit{Cybersecurity Knowledge and Skills Taught in Capture the Flag Challenges}, Elsevier Computers \& Security, 2020, ISSN 0167-4048, DOI: \texttt{10.1016/j.cose.2020.102154}, URL: \url{https://www.sciencedirect.com/science/article/pii/S0167404820304272}}
\newcommand\copyrightnotice{%
\begin{tikzpicture}[remember picture,overlay]
\node[anchor=north,yshift=-24pt] at (current page.north) {\fbox{\parbox{\dimexpr\textwidth-\fboxsep-\fboxrule\relax}{\copyrighttext}}};
\end{tikzpicture}%
}

\begin{document}

\begin{frontmatter}

\copyrightnotice

\title{Cybersecurity Knowledge and Skills Taught in Capture the Flag Challenges}

\author{Valdemar Švábenský (corresponding author)}
\ead{svabensky@ics.muni.cz}
\address{Masaryk University, Brno, Czech Republic}

\author{Pavel Čeleda}
\ead{celeda@ics.muni.cz}
\address{Masaryk University, Brno, Czech Republic}

\author{Jan Vykopal}
\ead{vykopal@ics.muni.cz}
\address{Masaryk University, Brno, Czech Republic}

\author{Silvia Brišáková}
\ead{brisakova@mail.muni.cz}
\address{Masaryk University, Brno, Czech Republic}

\begin{abstract}
Capture the Flag challenges are a popular form of cybersecurity education, where students solve hands-on tasks in an informal, game-like setting. The tasks feature diverse assignments, such as exploiting websites, cracking passwords, and breaching unsecured networks. However, it is unclear how the skills practiced by these challenges match formal cybersecurity curricula defined by security experts. We explain the significance of Capture the Flag challenges in cybersecurity training and analyze their 15,963 textual solutions collected since 2012. Based on keywords in the solutions, we map them to well-established ACM/IEEE curricular guidelines to understand which skills the challenges teach. We study the distribution of cybersecurity topics, their variance in different challenge formats, and their development over the past years. The analysis showed the prominence of technical knowledge about cryptography and network security, but human aspects, such as social engineering and cybersecurity awareness, are neglected. We discuss the implications of these results and relate them to contemporary literature. Our results indicate that future Capture the Flag challenges should include non-technical aspects to address the current advanced cyber threats and attract a broader audience to cybersecurity.
\end{abstract}


\end{frontmatter}


\section{Introduction}
\label{sec:intro}

Training security professionals is a slow but steady solution to the global cybersecurity workforce gap~\cite{isc2}. Educational institutions, computing societies, government organizations, and private companies are aware of this situation and introduce new curricula, study programs, and courses. Cybersecurity is an integral part of ACM/IEEE Computing Curricula 2020 (CC2020)~\cite{cc2020}, and specialized cybersecurity curricula, such as CSEC2017~\cite{curricula2017}, have been emerging in recent years.

Along with formal education, an increasingly popular method of practicing cybersecurity skills is via informal \textit{Capture the Flag} (CTF) games and competitions. In these events, small teams of participants exercise their cybersecurity skills by solving various tasks in an online learning environment. CTF tasks, called \textit{challenges}, feature diverse assignments from exploiting websites, through cracking passwords, to breaching unsecured networks. A successful solution of a challenge yields a text string called a \textit{flag} that is submitted online to prove reaching the solution.

CTF originated among cybersecurity enthusiasts at a hacker conference DEF CON in 1996~\cite{defcon}. However, CTF is no longer the niche of exclusive hacker groups. This educational game format quickly gained popularity, and now, teachers across the world are using it to complement education. CTF has been used successfully in university classes~\cite{vykopal2020, mirkovic2014} and in undergraduate security competitions~\cite{vigna2014, backman2016}. Even tech giants like Google and Facebook host CTFs~\cite{googlectf, facebookctf} that attract hundreds of attendees every year. Unlike traditional teaching formats, such as lectures and homework assignments, CTFs are more casual and often include competitive or game elements. However, because of their informality, it is unclear how they fit into cybersecurity curricula.

CTF participants publish their solutions to the challenges online. They do it to demonstrate solving the tasks and to share their knowledge with others. The solutions, called \textit{writeups}, are useful mainly in two ways. First, they are a learning resource that describes how the challenge was solved, which can prove useful in future CTFs and allow others to discover new solutions. Second, the writeups inspire CTF creators since they provide insight into the challenge assignment, even if the assignment is no longer available. We will investigate a third possible yet unexplored use of writeups. In our research, we regard them as a dataset and mine information about cybersecurity topics from them.

\subsection{Goals of This Paper}
\label{subsec:goal}

By analyzing the content of writeups, we examine how the informal CTF challenges map to formal CSEC2017 curricular guidelines defined by security experts. We seek to uncover the breadth of the cybersecurity topics that CTF can teach to enhance education and training. Specifically, we pose the following three research questions.
\begin{enumerate}
    \item \textit{What is the distribution of cybersecurity topics in CTF challenges?}
    \item \textit{How does the distribution of topics differ between various CTF formats?}
    \item \textit{How has the distribution of topics evolved over the past decade?}
\end{enumerate}
The first question explores dominant, typical, and underrepresented cybersecurity topics within the analyzed writeups. The second question divides the writeups according to the two most popular CTF formats and compares them, allowing educators to choose a suitable format for their learning activities. The third question splits the writeups based on the year of the corresponding CTF event and searches for trends over the years.

\subsection{Contributions of This Paper}

Online CTF challenges feature practical assignments, scale to hundreds of students, and include game elements. They are suitable for secondary, tertiary, professional, and extracurricular education. This paper aims to support their further transfer into the practice of teaching and learning security. Answering the research questions will be valuable for various stakeholders.
\begin{itemize}
    \item \textit{Cybersecurity experts} will know which cybersecurity skills they or their team members can practice via CTF.
    \item \textit{Educational managers} and \textit{curricular designers} can see how informal education via CTF helps fulfill formal cybersecurity learning outcomes. Moreover, at a higher level, they can consider which cybersecurity topics can be supplemented by CTF in their study programs.
    \item \textit{Teachers} and \textit{CTF content creators} may focus on more common cybersecurity knowledge to help students interested in CTF. Alternatively, they can teach the fields uncovered by CTF.
    \item \textit{Students} and \textit{CTF participants} can better understand the content of previous challenges and prepare for future challenges.
\end{itemize}

\subsection{Paper Structure}

This paper is organized into seven sections.
\Cref{sec:background} explains the key terms to familiarize the readers with CTF challenges, writeups, and cybersecurity curricula.
\Cref{sec:related-work} describes primary and secondary studies related to writeups, curricular design, and educational text analysis.
\Cref{sec:methods} details our methods for the collection and analysis of writeup data.
\Cref{sec:results} presents the findings and answers the three research questions.
\Cref{sec:lessons} offers practical insights and lessons learned from this research.
Finally, \Cref{sec:conclusion} concludes and summarizes our contributions.

\section{Background and Terminology}
\label{sec:background}

This section defines the key terms used throughout the paper: Capture the Flag in \Cref{subsec:CTF}, writeups and their web catalogs in \Cref{subsec:CTFtime}, and Cybersecurity Curricular Guidelines in \Cref{subsec:curricula}.

\subsection{Capture the Flag and Its Formats}
\label{subsec:CTF}

The term Capture The Flag originally refers to an outdoor game for two teams. Each team must simultaneously defend a (physical) flag in their base and steal the other team's flag. Since the 1990s, this playground has also moved to cyberspace. In cybersecurity, the term CTF denotes a broad spectrum of events with different scope and format. These include online competitions in attacking only, educational games with the instructional support of learners, or games played just for entertainment at popularization events.

CTF can be hosted on various technical platforms~\cite{Kucek2020}. The infrastructure is usually online, making CTF suitable for distance and blended learning. In fact, most CTFs are held remotely, although the participants can also meet on a physical location (see \Cref{table:ctf-event-count}). The most common formats of CTF are \textit{\jeopardy} and \textit{\attackdefense}.

In a \jeopardy\ CTF, the participants choose from challenges divided into categories, such as cryptography, reverse engineering, or gaining ownership of a service (\textit{pwn} in hacker jargon). Each challenge is usually worth a different number of points, imitating the format of the famous television show \textit{Jeopardy!}. The participants solve the challenges locally at their computers or interact with a remote server. An example task is that the participants receive a binary file containing an encrypted flag, which they have to recover.

In an \attackdefense\ CTF, each team of participants controls and maintains an identical instance of a computer network, whose hosts run vulnerable services. The goal is to patch the services and protect the network assets while exploiting the vulnerabilities in the services of other teams. The scoring is based on a combination of successful exploits and defensive countermeasures while maintaining the services' availability.

\subsection{CTF Web Catalogs}
\label{subsec:CTFtime}

\begin{figure}[!ht]
  \centering
  \includegraphics[width=\textwidth]{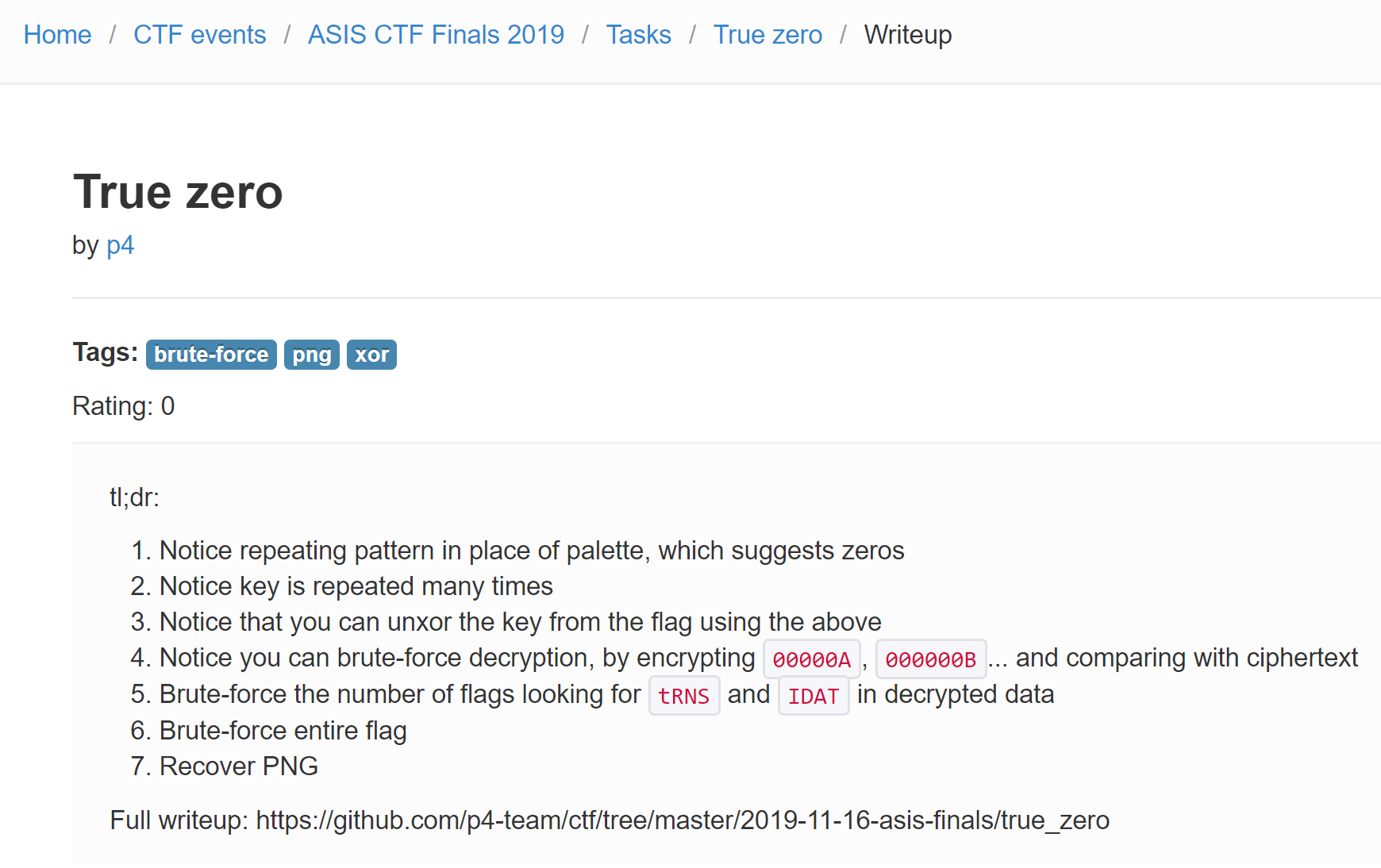}
  \caption{Example of a writeup. Source: \texttt{https://ctftime.org/writeup/17308}.}
  \label{figure:writeup-screenshot}
\end{figure}

CTF participants publish their writeups on different websites, such as GitHub, YouTube, or personal blogs. We focus on CTFtime.org~\cite{ctftime.org}, which is the most prominent web portal about CTF. Since its foundation in 2012, this community-run project has been collecting information about past CTF events, planned future events, team rankings, and challenge writeups. \Cref{figure:writeup-screenshot} shows an example of a writeup (a step-by-step solution) posted on the CTFtime website. It refers to the \jeopardy\ challenge from \Cref{subsec:CTF}.

Apart from \jeopardy\ and \attackdefense, CTFtime lists a third format of CTF: Hack Quest. However, this term is rarely used~\cite{Gondree2016Talking}, and according to the available information, we did not find any difference from \jeopardy. Therefore, we consider these two terms a synonym.

\Cref{table:ctf-event-count} shows that the number of CTFs listed at CTFtime increases each year. This growing popularity indicates that more people participate in CTF events. As a result, more data are generated that can be analyzed. Jeopardy is undoubtedly the dominant format, followed by \attackdefense. Moreover, approximately two-thirds of CTFs are available remotely. The rest takes place at a physical location (typically the finals). Remote CTFs are more accessible to a wide range of participants.

\newcolumntype{R}[1]{>{\raggedleft\let\newline\\\arraybackslash\hspace{0pt}}m{#1}}

\begin{table}[!ht]
\centering
\small
\rowcolors{2}{gray!10}{white}
\renewcommand*{\arraystretch}{1.25}
\begin{tabular}{|c||r|R{1.7cm}|R{1.35cm}||R{1.7cm}|R{1.7cm}||r|}
    \hline
    \rowcolor{gray!20}
     & \multicolumn{3}{|c||}{\textit{Division by Game Format}} & \multicolumn{2}{c||}{\textit{Division by Location}} & \\
    \textbf{Year} & \textbf{Jeopardy} & \centering\textbf{Attack-Defense} & \centering\textbf{Hack Quest} & \centering\textbf{Remote} & \centering\textbf{On-site} & \textbf{Total} \\
    \hline
    2012 &  23 & 10 & 2 &  19 & 16 &  35 \\
    2013 &  41 & 13 & 1 &  35 & 20 &  55 \\
    2014 &  49 &  8 & 1 &  36 & 22 &  58 \\
    2015 &  65 & 12 & 2 &  48 & 31 &  79 \\
    2016 &  90 & 14 & 3 &  67 & 40 & 107 \\
    2017 & 125 & 14 & 2 & 102 & 39 & 141 \\
    2018 & 136 & 16 & 1 & 102 & 51 & 153 \\
    2019 & 175 & 20 & 3 & 145 & 53 & 198 \\
    2020 & 126 & 13 & 4 & 130 & 13 & 143 \\
    \hline
    Total & 830 (86\%) & 120 (12\%) & 19 (2\%) & 684 (71\%) & 285 (29\%) & 969 \\
    \hline
\end{tabular}
\caption{The numbers of all CTF events posted on CTFtime.org~\cite{ctftime.org} from January 1, 2012, to October 9, 2020, divided according to the game format and location. Since the year 2020 is incomplete, it has fewer events so far. The small number of on-site events in 2020 is probably due to worldwide COVID-19 restrictions.}
\label{table:ctf-event-count}
\end{table}

\subsection{Cybersecurity Curricular Guidelines}
\label{subsec:curricula}

Curricula are formal documents that describe the knowledge and skills taught at an educational institution. Prominent curricula in computing, such as the ACM/IEEE 2013 computing curricula~\cite{Sahami2013computer} and CC2020~\cite{cc2020}, include only broad cybersecurity topics. That is why specialized cybersecurity curricula started to emerge (see~\cite{Parrish2018, Mouheb2019} for a detailed overview). Among these, we chose the \textit{Cybersecurity Curricular Guidelines} (CSEC2017) developed by The Joint Task Force on Cybersecurity Education~\cite{curricula2017} because they are widely established in the field.

CSEC2017 defines eight \textit{Knowledge Areas} (KAs) in cybersecurity. Each KA encompasses different skills and knowledge.
\begin{enumerate}[itemsep=0pt]
    \item \textit{Data security} includes cryptography, forensics, data integrity, and authentication.
    \item \textit{Software security} focuses on secure programming, testing, and other aspects of software development.
    \item \textit{Component security} deals with the security of components integrated into larger systems, including their design and reverse engineering.
    \item \textit{Connection security} encompasses network services, defense, and attacks.
    \item \textit{System security} aims at securing a system as a whole, including access control and penetration testing.
    \item \textit{Human security} is about protecting individuals' identity, data, and privacy. It includes social engineering and cybersecurity awareness.
    \item \textit{Organizational security} governs risk management, security policies, and incident handling at the level of organizations.
    \item \textit{Societal security} examines cybersecurity at the national or global level. It concerns cybercrime, cyber law, and governmental policies.
\end{enumerate}

\begin{figure}[!ht]
  \centering
  \includegraphics[width=\textwidth]{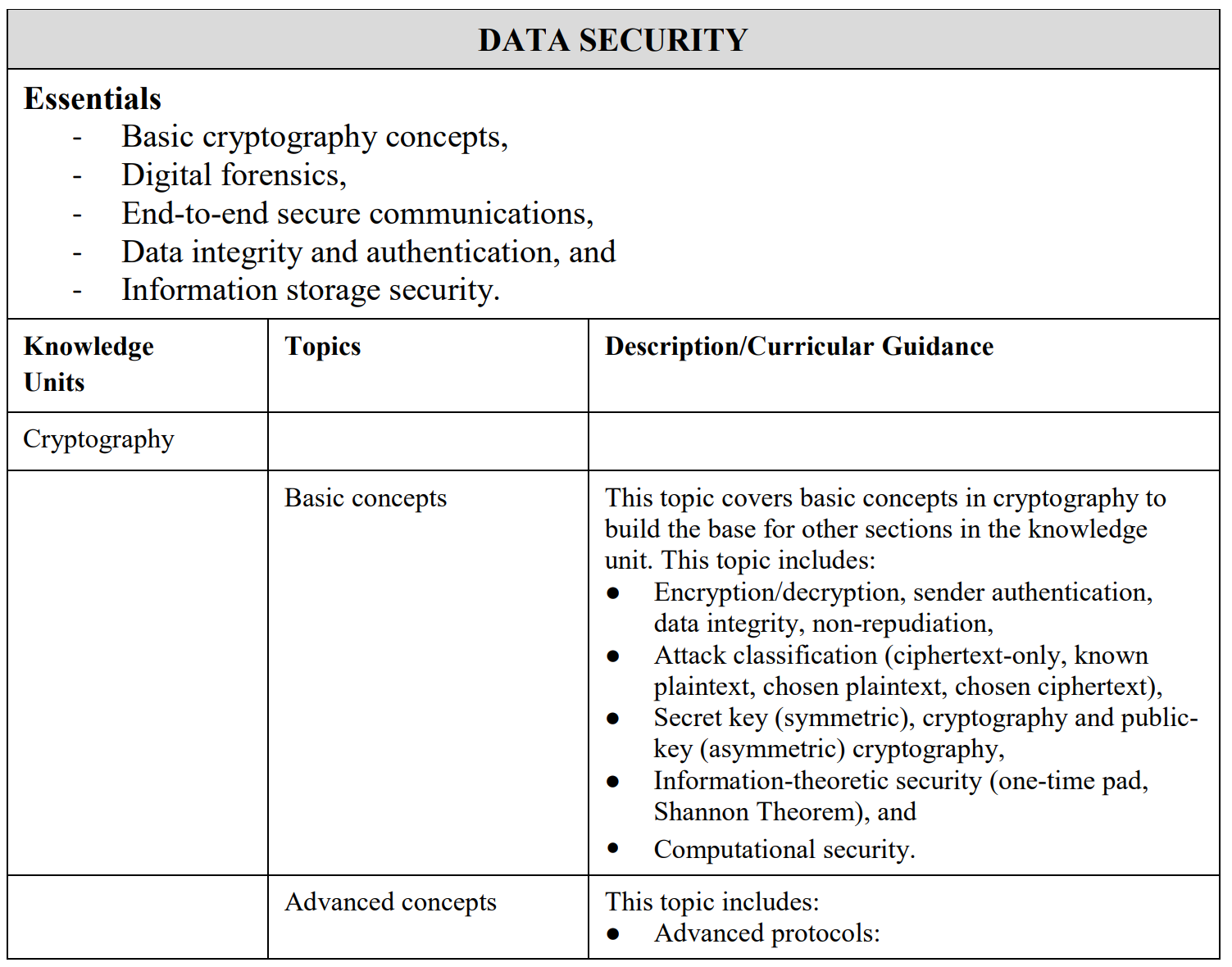}
  \caption{Excerpt from the CSEC2017 curricular guidelines~\cite[Chapter 4, page 24]{curricula2017}.}
  \label{figure:curricula-screenshot}
\end{figure}

Each KA is divided into \textit{Knowledge Units} (KUs), which are further split into \textit{Knowledge Topics}. (Although CSEC2017 uses \textit{Topic}, we changed it to \textit{Knowledge Topic} to introduce the abbreviation KT.) \Cref{figure:curricula-screenshot} provides an example of Data security KA. Its first KU, Cryptography, forms the first column of the table. The second column contains the names of the subordinate KTs. The third column describes the knowledge and skills that belong to the KTs. Overall, there are 8 KAs, 55 KUs, and 287 KTs. This paper identifies the distribution of the KAs and KUs in the writeups of CTF challenges.

\section{Related Work}
\label{sec:related-work}

This section presents the related publications and explains how this research differs from state of the art.

\subsection{Analysis of CTF Writeups}
\label{subsec:related-work-ctf}

The closest research publication to this paper is by Burns et al.~\cite{Burns2017Analysis}. They analyzed CTF writeups to find essential skills and knowledge to study when preparing to participate in CTF events. The work focused on 160 events with about 3,600 solutions posted on GitHub in the years 2011--2016. They analyzed the data to develop challenges for beginners, which are grouped into six categories: \textit{crypto}, \textit{web}, \textit{reverse}, \textit{forensic}, \textit{pwn}, and \textit{misc}. Although the article did not specify the exact details of analysis methods applied to CTF writeups, the authors likely read and classified the writeups manually. We aim to automate the process to achieve the results faster and more reliably by reducing human errors. Moreover, we map the results to formally defined Knowledge Areas and Units of a cybersecurity curriculum.

In a paper by Chothia et al.~\cite{chothia2015}, students of a cybersecurity course submitted writeups to CTF challenges they solved. The researchers manually graded these writeups and examined the correlations of the resulting grades with the number of submitted flags. Moreover, they considered the writeups as an indicator of whether students had understood the learning content. Similarly, Schreuders et al.~\cite{schreuders2017} and Leune et al.~\cite{Leune2017} instructed their students to submit writeups of CTF challenges. However, the two papers did not mention further analysis of the writeups. Overall, there was no published attempt to map CTF writeups to a cybersecurity curriculum.

\subsection{Analysis of Cybersecurity Skills and Curricula}
\label{subsec:related-work-cybered}

Cabaj et al.~\cite{Cabaj2018Cybersecurity} analyzed cybersecurity topics in 21 master's degree programs at top-ranking universities worldwide. They informed how the programs cover cybersecurity topics and how the topics are distributed among the taught courses. For the topic analysis, they chose ACM/IEEE 2013 computing curricula~\cite{Sahami2013computer} complemented with the CSEC2017~\cite{curricula2017}. Their results revealed the increasing importance of non-technical cybersecurity areas: \textit{Human}, \textit{Organizational}, and \textit{Societal security}. In related work, Hallett et al.~\cite{Hallett2018} mapped the content of various cybersecurity curricula onto the knowledge areas from the United Kingdom's Cybersecurity Body of Knowledge~\cite{cybok}.

In a technical report from a European Union cybersecurity project~\cite{CS4EUreport}, 96 universities were surveyed about which KAs from CSEC2017 they cover. They identified that \textit{Data} and \textit{Connection security} are covered the most, while \textit{Organizational security} is covered the least.

Carroll~\cite{Carroll2018Offensive} investigated the skills required in cyber warfare, specifically for developing a workforce in offensive and defensive cyberspace operations. He surveyed 23 cyberspace professionals from the military, civilian, and private sector about their core knowledge, skills, and abilities. As a result, he recommends actions to improve the preparation for cyberspace operations.

Jones et al.~\cite{Jones2018Core} examined the essential knowledge, skills, and abilities for cybersecurity jobs. They surveyed 44 cybersecurity professionals attending the major hacker conferences Black Hat and DEF CON. As a result, they suggest prioritizing knowledge about networks, vulnerabilities, programming, and interpersonal communication. In related work, Haqaf and Koyuncu~\cite{Haqaf2018Understanding} used the Delphi method~\cite{linstone1975delphi} (a structured communication technique) to discover the key skills for information security managers. Finally, Brooks et al.~\cite{Brooks2018Information} analyzed IT security job advertisements to determine the skills that employers are interested in.

In our previous work~\cite{Svabensky2020}, we performed a literature review of 71 cybersecurity education papers. As a part of the review, we mapped the content of the papers to the CSEC2017. The dominant KA was \textit{Data security}. Other technical KAs, \textit{Software}, \textit{Connection}, and \textit{System security} were strongly present too. Nevertheless, \textit{Human}, \textit{Organizational}, and \textit{Societal security} were not neglected either. \textit{Component security} was the least common.

\subsection{Analysis of Topics in Other Textual Data to Support Education}
\label{subsec:related-work-topic-modeling}

Latent Dirichlet allocation (LDA) is a method for probabilistic topic modeling~\cite{Blei2012Probabilistic}. Given a set of text documents, LDA discovers their underlying topics, which are probability distributions over the words (terms) in the documents. LDA is a commonly used method that addresses the limitations of probabilistic latent semantic indexing~\cite{Hofmann1999Probabilistic}. Marçal et al.~\cite{Marcal2020} used LDA to identify computing topics in questions asked on Stack Overflow. Similarly, Rouly et al.~\cite{Rouly2015} used LDA to analyze course descriptions in course catalogs of different universities.

Nadeem et al.~\cite{Nadeem2015method} developed a method for recommending relevant reading materials to software developers based on vulnerabilities in their code. They statically analyzed program source code to discover the vulnerabilities. Then, they computed the cosine similarity between the description of the vulnerabilities and public articles in the Common Weakness Enumeration (CWE) database. To model the articles, they used a standard approach of Term Frequency -- Inverse Document Frequency (TF--IDF)~\cite[p. 302]{handbook-edm2010}. It assigns weights to the words in text based on their frequency and importance.

\section{Methods}
\label{sec:methods}

This section explains the methods we chose to answer the research questions posed in \Cref{subsec:goal}.

\subsection{Extracting Cybersecurity Keywords from the CSEC2017 Guidelines}
\label{subsec:keywords}

To identify cybersecurity topics covered by CTFs, we searched for cybersecurity keywords in the writeups. We began by extracting these keywords from the CSEC2017~\cite{curricula2017}. For each of the eight tables with Knowledge Areas (see the example in \Cref{figure:curricula-screenshot}), one author manually extracted keywords from the third column, \enquote{Description/Curricular Guidance}. Another author revised the extraction. Then, we repeated the process to minimize errors and ensure the inclusion of all relevant keywords.

We gathered 1,623 keywords and organized them in a JavaScript Object Notation (JSON) file~\cite{Severance2012}. Its excerpt is shown in \Cref{figure:keywords-json}, and its structure follows~\Cref{figure:curricula-screenshot}. KAs are JSON objects that contain other objects, KUs. Similarly, KUs contain an array of KT objects, which then contain individual keywords. To verify the file's correctness and syntax, we wrote a JSON validation schema~\cite{droettboom2015understanding} for it.

\begin{figure}[!ht]
\begin{verbatim}
    "Knowledge Area": "Data Security",
    "Knowledge Units": [
        {
            "name": "Cryptography",
            "Knowledge Topics": [
                {
                    "name": "Basic Concepts",
                    "keywords": [
                        "encryption",
                        "decryption",
                        "sender authentication",
                        ...
\end{verbatim}
\caption{Excerpt from the JSON file with the cybersecurity keywords we searched for in the writeups. Notice how the content corresponds to the curricula excerpt in \Cref{figure:curricula-screenshot}.}
\label{figure:keywords-json}
\end{figure}

\subsection{Downloading the Writeups}
\label{subsec:downloading-writeups}

More than 969 CTF events with a total of \numtotaltasks\ challenges (tasks) and \numtotalwriteups\ writeups have been posted on CTFtime since its foundation in 2012. We focused on the writeups of events that took place from January 1, 2012, to October 9, 2020, since 2012 is the first year to contain any writeups. To download the content of the writeups, we used the Python \texttt{requests} library~\cite{pythonrequests} to implement the algorithm sketched in \Cref{figure:algorithm-download}.

\begin{figure}[!ht]
\begin{verbatim}
    for each event posted on CTFtime.org:
        for each task in the event:
            for each writeup of the task:
                download the text of the writeup
\end{verbatim}
\caption{Pseudocode for downloading the writeups.}
\label{figure:algorithm-download}
\end{figure}

Downloading the writeup text differed based on the content of the writeup. There were three possibilities:
\begin{enumerate}
    \item The writeup text was present directly on the CTFtime webpage. In this case, we simply scraped it.
    \item The writeup included only a link to an external website. If the website was GitHub, which was the most common case, we followed the link. Then, if the text contained at least one keyword, we scraped the raw file content. For other websites, such as the authors' blogs, we ignored the link. The reason was that each website had a different structure, so automating the download would be time-consuming.
    \item The writeup included a combination of the text and an external link. If the external link was not on GitHub, we scraped the text as in option~1. If the external link was on GitHub, we scraped both the CTFtime text (option 1) and the GitHub text (option 2). Then, we counted the cybersecurity keywords in both files and kept the file with more keywords since we considered it more representative. Often, CTFtime included a sketch of the writeup (its subset) and GitHub its full text.
\end{enumerate}

Then, we performed data cleaning, such as removing HTML tags and links to external websites. After the cleaning, we discarded writeups shorter than two characters because our shortest keyword was two characters long. The remaining writeups were categorized by year and format and saved as a text file. Altogether, these files act as the input for the analysis (see \Cref{subsec:data-analysis}).

\Cref{table:downloaded-writeups} shows the resulting number of downloaded writeups categorized by years and format. All the events we worked with are \jeopardy\ or \attackdefense. Although 19 events were Hack Quests, none contained any writeup, so we excluded them.

\begin{table}[!ht]
\centering
\small
\rowcolors{2}{gray!10}{white}
\renewcommand*{\arraystretch}{1.25}
\begin{tabular}{|c|r|r|r|}
    \hline
    \rowcolor{gray!20}
    \textbf{Year} & \textbf{Jeopardy}\hspace*{1.5mm} & \textbf{Attack-Defense} & \textbf{Total} \\
    \hline
    2012 &    90 &  9 &    99 \\
    2013 &   145 &  3 &   148 \\
    2014 &   118 &  1 &   119 \\
    2015 &   419 &  0 &   419 \\
    2016 & 1,927 &  4 & 1,931 \\
    2017 & 2,499 & 15 & 2,514 \\
    2018 & 3,919 & 21 & 3,940 \\
    2019 & 3,153 & 17 & 3,170 \\
    2020 & 3,609 & 14 & 3,623 \\
    \hline
    Total & 15,879 (99\%) & 84 (1\%) & 15,963 \\
    \hline
\end{tabular}
\caption{The numbers of downloaded and subsequently analyzed writeups posted on CTFtime.org~\cite{ctftime.org} from January 1, 2012, to October 9, 2020.}
\label{table:downloaded-writeups}
\end{table}

Several factors caused the difference between the total number of writeups (\numtotalwriteups) and the downloaded writeups (\numdownloadedwriteups). The most common one was that the writeup was linked on an external website or written in a PDF, which we did not parse. In rare cases, the writeup was empty, deleted by the author, or did not pass through the data cleaning process.

\subsection{Analyzing the Downloaded Writeups to Identify the Keywords}
\label{subsec:data-analysis}

\begin{figure}[!ht]
\centering
\begin{tikzpicture}[node distance = 0.8cm, thick]

\node (1) {\shortstack{{\Large\faFileTextO}\\[2mm]CSEC2017\\curricula}};
\node (2) [right=of 1] {\shortstack{{\Large\faCogs}\\[2mm]Manual\\extraction}};
\node (3) [right=of 2] {\shortstack{{\Large\faFolderOpen}\\[2mm]1,623\\keywords}};
\node (4) [below=of 1] {\shortstack{{\Large\faGlobe}\\[2mm]CTFtime\\website}};
\node (5) [right=of 4] {\shortstack{{\Large\faFileCodeO}\\[2mm]Downloading\\script}};
\node (6) [right=of 5] {\shortstack{{\Large\faFolderOpen}\\[2mm]\numdownloadedwriteups \\writeups}};
\node (7) [below right=-0.5cm and 0.75cm of 3] {\shortstack{{\Large\faFileCodeO}\\[2mm]Analysis\\script}};
\node (8) [right=of 7] {\shortstack{{\Large\faBarChart}\\[2mm]Keyword\\frequencies}};

\draw[->] (1) -- (2);
\draw[->] (2) -- (3);
\draw[->] (4) -- (5);
\draw[->] (5) -- (6);
\draw[->] (3) -- (7);
\draw[->] (6) -- (7);
\draw[->] (3) -- (5);
\draw[->] (7) -- (8);
\end{tikzpicture}
\caption{Overview of the data collection and analysis pipeline.}
\label{figure:analysis-overview}
\end{figure}
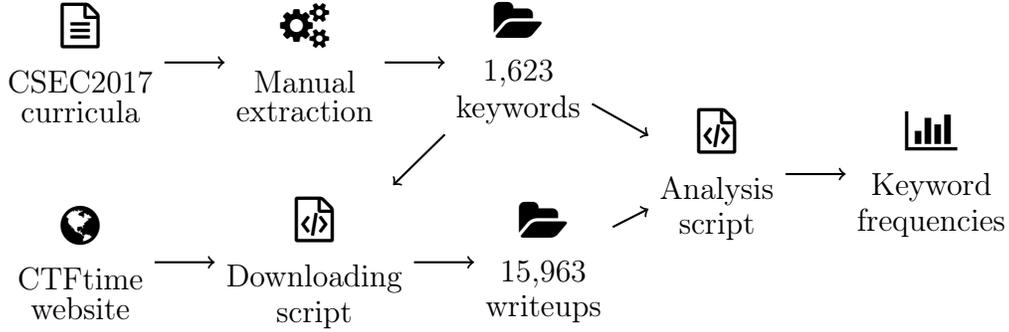

\Cref{figure:analysis-overview} shows an overview of all entities that take part in the analysis. The analysis script takes two inputs: the keywords file and the downloaded writeups. Each writeup is represented using a \textit{Bag of words} model~\cite{handbook-la2017}, so the order of the words is disregarded, and we count the keywords in each writeup. Formally, we define the analysis as follows.

\subsubsection{Input for the Analysis Script}
The script for analyzing the writeups has two inputs:
\begin{itemize}
    \item $K = \{k_{1},\dots,k_{n}\}$, a set of $N$ keywords. We defined $N = 1{,}623$ keywords. Each $k_{i}$ belongs to exactly one KT, which belongs to exactly one KU, which belongs to exactly one KA.
    \item $W = \{w_{1},\dots,w_{m}\}$, a set of $M$ writeups. Overall, $M = 15{,}963$. $W$ is partitioned into subsets for the second and third research questions.
\end{itemize}

\subsubsection{Counting the Keyword Frequencies}
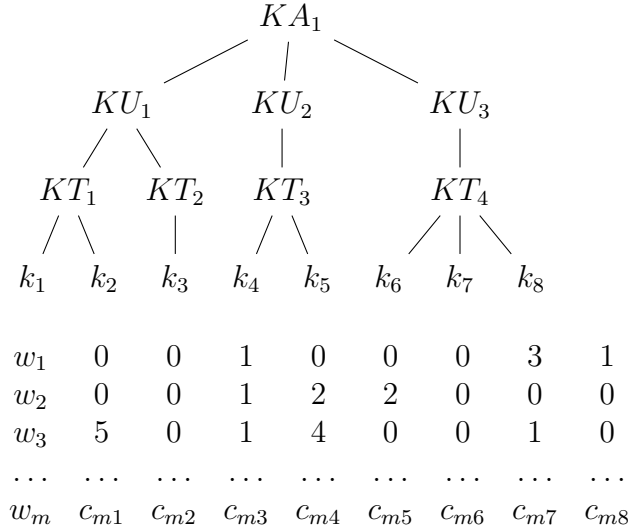
\begin{figure}[!ht]
\begin{center}
\hspace{-10mm}%
\begin{forest}
baseline
[$KA_{1}$
    [$KU_{1}$
        [$KT_{1}$,l sep=5mm,
            [$k_{1}$][$k_{2}$]
        ],[$KT_{2}$[$k_{3}$]]
    ], s sep=4mm, 
  [$KU_{2}$
    [$KT_{3}$,
        [$k_{4}$] [$k_{5}$]
        ]
    ],
  [$KU_{3}$
        [$KT_{4}$, [$k_{6}$] [$k_{7}$], [$k_{8}$]
    ]
  ]
]
\end{forest}

$$
\begin{matrix}
\arraycolsep=3.9pt\def\arraystretch{20.2}
w_{1} & 0 & 0 & 1 & 0 & 0 & 0 & 3 & 1 \\
w_{2} & 0 & 0 & 1 & 2 & 2 & 0 & 0 & 0 \\
w_{3} & 5 & 0 & 1 & 4 & 0 & 0 & 1 & 0 \\
\dots & \dots & \dots & \dots & \dots & \dots & \dots & \dots & \dots \\
w_{m} & c_{m1} & c_{m2} & c_{m3} & c_{m4} & c_{m5} & c_{m6} & c_{m7} & c_{m8}
\end{matrix}
$$
\end{center}
\caption{Example of counting the keyword frequencies in the writeups.}
\label{figure:analysis-example}
\end{figure}

Given $W$ and $K$, the goal is to compute $C = (c_{ij})$, an $M \times N$ matrix, where each $c_{ij}$ is the count of occurrences of keyword $k_{j}$ in writeup $w_{i}$. The value $c_{ij}$ is further referred to as \textit{Term Frequency} (TF). \Cref{figure:analysis-example} shows an illustrative example of counting the keywords in the writeups.

Note that the matrix will be sparse; it will contain a zero value for each keyword that does not occur in the writeup $w_{i}$. Since each writeup explains the solution to a single challenge, it is extremely unlikely that many of the 1,623 keywords would be present in a single writeup.

When counting the keywords in writeups, the matching is always case insensitive because the writeups are often informal and written in lowercase. Moreover, two additional rules are applied. If the keyword is an abbreviation, such as LAN (for Local Area Network), we seek its exact match. However, if the keyword is not an abbreviation, we seek only its stem~\cite{lovins1968stemming}. For example, if our defined keyword is \enquote{encryption}, and the writeup contains the word \enquote{encrypting}, we consider it a match.

\subsubsection{Normalizing the Term Frequencies}
The writeups have different lengths. Naturally, it is more likely for longer writeups to include more keywords. Therefore, to eliminate the bias of longer writeups over the shorter ones, we need to normalize the TF values. We could normalize by dividing each TF by the length of the writeup, but this would give us impractically small numbers. Therefore, we decided to divide all TF values by their sum within each row. For example:

$$w_{1}\text{: } [0, 0, 1, 0, 0, 0, 3, 1] \xrightarrow[\text{}]{\text{normalization}} [0, 0, \tfrac{1}{5}, 0, 0, 0, \tfrac{3}{5}, \tfrac{1}{5}]$$

This normalization yields a value we call the \textit{Normalized Term Frequency} (NTF), which has multiple benefits. First, it maps all the values in the matrix $C$ to a common range $[0,1]$. Second, it preserves the relative differences between the original TF values. Third, the NTF values sum to 1 for each writeup $w_{i}$, and so can be easily represented by percentages.

\subsubsection{Assigning the Writeups to KUs and KAs}
The process of assigning a writeup $w_{i}$ to a KU is as follows. Each KU is assigned the sum of NTFs of its respective keywords. For example, suppose that $KU_{1}$ contains keywords $k_{1}$, $k_{2}$, $k_{3}$. Then $c_{i1}$ + $c_{i2}$ + $c_{i3}$ is assigned to $KU_{1}$. Returning to the example matrix in \Cref{figure:analysis-example}, given only the writeup $w_1$, $KU_{1}$ would receive the value $\frac{1}{5}$, $KU_{2}$ the value $0$, and $KU_{3}$ $\frac{4}{5}$.

This calculation is applied to all writeups and all 55 KUs. Afterward, the assigned values are normalized by dividing them by $M$, the total number of writeups. We do this to achieve the same benefits as for the normalization above. Finally, the process is analogous for grouping KUs into KAs. The output of the analysis is the distribution of KUs/KAs in the writeups $W$.

We publish the analysis script with the supplementary materials for this paper~\cite{dataset}. Its basis was created in a previous project~\cite{Brisakova2020thesis} and updated for this work. See also \Cref{sec:conclusion} for details.

\section{Results and Discussion}
\label{sec:results}

This section answers the three research questions (RQ) about the distribution of cybersecurity topics overall, in the two CTF formats, and throughout the years 2012--2020.

\subsection{RQ1: Distribution of Knowledge Areas and Units in CTF Writeups}
\label{subsec:results-rq1}

We now answer the first research question by looking into the distribution of cybersecurity topics in the \numdownloadedwriteups\ analyzed writeups. In total, we identified 232,160 keyword matches, corresponding to about 14.5 keywords per writeup on average. Out of the 1,623 keywords, 1,012 were not found in any writeup. The ten most common keywords were: \textit{log}, \textit{password}, \textit{exploit}, \textit{encrypt},  \textit{class}, \textit{pwn}, \textit{http}, \textit{decrypt}, \textit{crypto}, and \textit{reverse}.

\subsubsection{Overall Distribution of Knowledge Areas}
\label{subsubsec:rq1-KA}

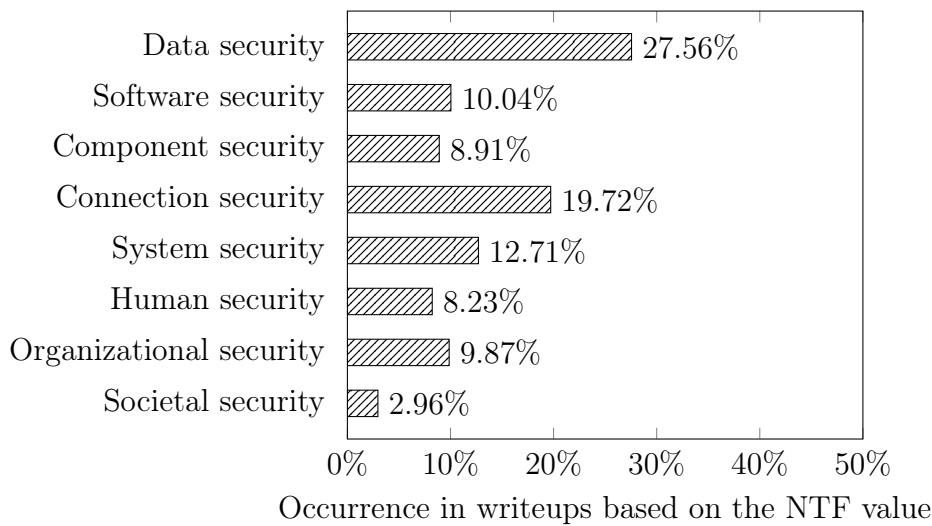
\begin{figure}[!ht]
\centering
\begin{tikzpicture}
  \begin{axis}[
    xbar,
    xmin = 0,
    xmax = 50,
    xlabel = {Occurrence in writeups based on the NTF value},
    symbolic y coords={Societal security, Organizational security, Human security, System security, Connection security, Component security, Software security, Data security},
    ytick = data,
    ytick style={opacity=0},
    xticklabel={\pgfmathprintnumber\tick\%},
    nodes near coords=\pgfmathprintnumber{\pgfplotspointmeta}\%,
  ]
  \addplot [fill=gray!50,pattern=north east lines] coordinates {(2.96,Societal security) (9.87,Organizational security) (8.23,Human security) (12.71,System security) (19.72,Connection security) (8.91,Component security) (10.04,Software security) (27.56,Data security)};
  \end{axis}
\end{tikzpicture}
\caption{Overall distribution of KAs in the downloaded writeups.}
\label{result:rq1-KAs}
\end{figure}

\Cref{result:rq1-KAs} shows the distribution of cybersecurity KAs in the writeups. We can see that the analyzed writeups incorporate each KA to some extent. The most prominent is \textit{Data security}: among all keyword matches, more than 27\% corresponded to it. The second place is taken by \textit{Connection security}, and \textit{System security} is the third.

These three KAs are above the average of 12\%. They are the most popular probably due to containing skills and knowledge that are suitable to test in CTF challenges. \textit{Data security} includes knowledge about cryptography, authentication, and secure communication. \textit{Connection security} comprises network services and defense. Finally, \textit{System security} involves penetration testing and multi-stage attacks.

Next, KAs \textit{Software, Organizational, Component}, and \textit{Human security} have similar values of around 8--10\%, which are below the average. The third least is \textit{Component security}, possibly because it often involves using physical devices. This requires the device to be physically present next to the participant, which is complicated and costly for remote CTFs.

The least frequent KA is \textit{Societal security}, with only 3\% value. This is not surprising since societal aspects of cybersecurity, such as privacy or cyber law, are not usually covered by CTF challenges. The creators of CTF primarily focus on technical knowledge.

\subsubsection{Overall Distribution of Knowledge Units}
\label{subsubsec:rq1-KU}

\Cref{table:results-rq1-KU} lists the top ten prominent KUs we identified, along with their key KT and a parent KA to provide context. \textit{Cryptography} is the most prominent KU, arguably because of dealing with ciphers. Tasks requiring a participant to decrypt data are popular, and may also be used as a subtask in various bigger tasks. Moreover, symmetric and asymmetric cryptography are often core content in cybersecurity courses.

\begin{table}[!ht]
\centering
\small
\rowcolors{2}{gray!10}{white}
\renewcommand*{\arraystretch}{1.25}
\begin{tabular}{|r|>{\raggedright}m{3.6cm}|>{\raggedright}m{3.6cm}|>{\raggedright}m{3.5cm}|}
    \hline
    \rowcolor{gray!20}
    \textbf{\%} & \textbf{Knowledge Unit} & \textbf{Key Topic} & \textbf{Parent Knowledge Area} \tabularnewline
    \hline
    14.9 & Cryptography & Encryption & Data security \tabularnewline
    8.8 & Component design & Reverse engineering & Component security \tabularnewline
    7.1 & Implementation & Secure programming & Software security \tabularnewline
    7.0 & System control & Penetration testing & System security \tabularnewline
    6.5 & Digital forensics & Artifact analysis & Data security \tabularnewline
    5.8 & Distributed systems architecture & HTTP(S), web attacks & Connection security \tabularnewline
    5.3 & Network services & Network protocols and attacks on them & Connection security \tabularnewline
    5.3 & Network implementations & TCP/IP, network attacks & Connection security \tabularnewline
    3.9 & Identity management & Authentication & Human security \tabularnewline
    3.8 & Business continuity, disaster recovery, and incident management & Incident response & Organizational security \tabularnewline
    \hline
\end{tabular}
\caption{Ten most prominent KUs overall sorted by the NTF value.}
\label{table:results-rq1-KU}
\end{table}

On the other hand, one of the 55 KUs was not present at all: \textit{Physical interfaces and connectors}. We anticipated this result since most CTFs are remote and rarely involve the hacking of physical devices. Other least prominent KUs are again mostly about physical devices or belong to the societal parts of cybersecurity, such as \textit{Cyber law}.

\subsection{RQ2: Distribution of Knowledge Areas and Units in Jeopardy and Attack-Defense CTF Formats}
\label{subsec:results-rq2}

This section answers the second research question. We compare the distribution of KAs and KUs in \jeopardy\ and \attackdefense\ formats separately.

\subsubsection{Distribution of Knowledge Areas in Jeopardy and Attack-Defense CTF}
\label{subsubsec:rq2-KA}

\begin{figure}[!ht]
\centering
\begin{tikzpicture}
  \begin{axis}[
    xbar,
    xmin = 0,
    xmax = 50,
    xlabel = {Occurrence in writeups based on the NTF value},
    symbolic y coords={Societal security, Organizational security, Human security, System security, Connection security, Component security, Software security, Data security},
    ytick = data,
    ytick style={opacity=0},
    xticklabel={\pgfmathprintnumber\tick\%},
    nodes near coords=\pgfmathprintnumber{\pgfplotspointmeta}\%,
    legend style={at={(1,0)}, anchor=south east, legend columns=1, draw=none},
    reverse legend,
    legend cell align={left},
    legend image code/.code={%
      \draw[#1] (0cm,-0.1cm) rectangle (0.5cm,0.1cm);
    }, 
    y=1cm
    ]
  \addplot [fill=white] coordinates {(2.17,Societal security) (11.38,Organizational security) (9.18,Human security) (10.96,System security) (32.68,Connection security) (2.08,Component security) (14.72,Software security) (16.83,Data security)};
  \addplot [fill=gray!50] coordinates {(2.96,Societal security) (9.86,Organizational security) (8.23,Human security) (12.72,System security) (19.66,Connection security) (8.94,Component security) (10.02,Software security) (27.61,Data security)};
  \legend{Attack-Defense, Jeopardy}
  \end{axis}
\end{tikzpicture}
\caption{Distribution of KAs in 15,879 \jeopardy\ and 84 \attackdefense\ writeups.}
\label{result:rq2-KAs}
\end{figure}
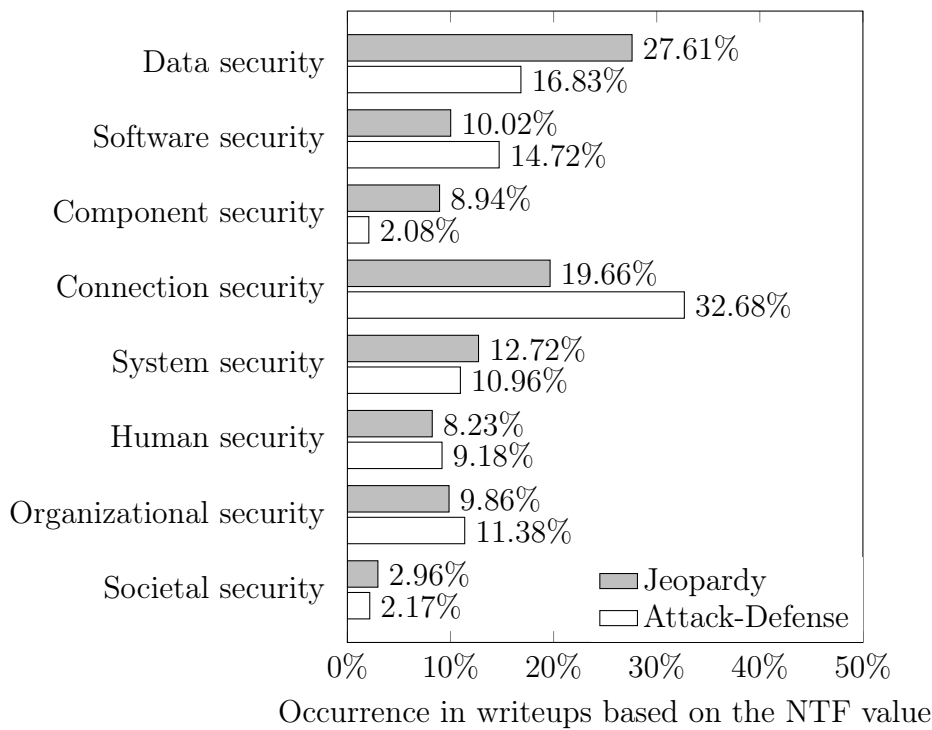

\Cref{result:rq2-KAs} shows that the distribution of \jeopardy\ writeups closely resembles the overall distribution shown in \Cref{result:rq1-KAs}. This is natural, since \jeopardy\ writeups constitute more than 99\% of the total (see \Cref{table:downloaded-writeups}). However, when looking at \attackdefense\ writeups alone, it is apparent that the distribution differs. The top three \jeopardy\ KAs are \textit{Data}, \textit{Connection}, and \textit{System security}. For \attackdefense, however, \textit{Connection security} dominates, followed by \textit{Data} and \textit{Software security}.

\textit{Connection security} is prominent in \attackdefense\ because this format heavily relies on networking skills. By definition, \attackdefense\ focuses on attacking systems of opposing teams via a network. Typically, this requires the participants to exploit network services, analyze traffic, or obtain credentials for a remote connection. On the other hand, \jeopardy\ often includes standalone cryptographic challenges, giving rise to \textit{Data security}.

\subsubsection{Distribution of Knowledge Units in Jeopardy and Attack-Defense CTF}
\label{subsubsec:rq2-KU}

Again, the distribution of KUs for the \jeopardy\ format alone follows the overall trends in \Cref{table:results-rq1-KU}. Because \attackdefense\ CTFs have few writeups, they had almost no impact on the most prominent KUs overall. Out of the top ten KUs in \jeopardy, three of them belong to KA Connection Security: \textit{Distributed systems architecture}, \textit{Network services}, and \textit{Network implementations}. Data Security contains \textit{Cryptography} and \textit{Digital forensics}, and the remaining KAs, excluding Societal security, have one representative each.

For the \attackdefense\ format, Connection security has three KUs again among the top ten (see \Cref{table:results-rq2-KU}). Data security contains \textit{Cryptography} and \textit{Digital forensics}. Software security contains \textit{Implementation} and \textit{Deployment and maintenance}. Again, the remaining KAs, excluding Component and Societal security, have one representative each.

\begin{table}[!ht]
\centering
\small
\rowcolors{2}{gray!10}{white}
\renewcommand*{\arraystretch}{1.25}
\begin{tabular}{|r|>{\raggedright}m{3.6cm}|>{\raggedright}m{3.8cm}|>{\raggedright}m{3.5cm}|}
    \hline
    \rowcolor{gray!20}
    \textbf{\%} & \textbf{Knowledge Unit} & \textbf{Key Topic} & \textbf{Parent Knowledge Area} \tabularnewline
    \hline
    12.3 & Network implementations & TCP/IP, network attacks & Connection security \tabularnewline
    10.1 & Distributed systems architecture & HTTP(S), web attacks & Connection security \tabularnewline
    8.9 & Implementation & Secure programming & Software security \tabularnewline
    8.3 & Digital forensics & Artifact analysis & Data security \tabularnewline
    8.2 & Network services & Network protocols and attacks on them & Connection security \tabularnewline
    7.2 & Business continuity, disaster recovery, and incident management & Incident response & Organizational security \tabularnewline
    6.7 & System control & Penetration testing & System security \tabularnewline
    5.7 & Social engineering & Deception & Human security \tabularnewline
    5.1 & Deployment and maintenance & Software configuration and patching & Software security \tabularnewline
    5.1 & Cryptography & Encryption & Data security \tabularnewline
    \hline
\end{tabular}
\caption{Ten most prominent KUs for the \attackdefense\ format sorted by the NTF value.}
\label{table:results-rq2-KU}
\end{table}

Regarding the least frequent KUs, we observed one KU not present in \jeopardy\ writeups, but 22 KUs were not identified in any of the \attackdefense\ writeups. An example in both formats is \textit{Physical Interfaces and Connectors}, and we add the full list in the supplementary materials~\cite{dataset}. These may inspire cybersecurity educators to design new types of CTF.

We statistically compared whether the KU distribution difference between \jeopardy\ and \attackdefense\ is significant. Before choosing a statistical test, we ran a normality test on the whole computed matrix $C$. We chose the Anderson–Darling test implemented in the Python \texttt{scipy} library~\cite{PythonAndersonDarling}, which strongly rejected the hypothesis of the data having a normal distribution.

Due to the skewed data distribution, we considered only non-parametric statistical tests. We chose the Mann-Whitney U test implemented in the Python \texttt{scipy} library~\cite{PythonMannWhitneyU}. Each of the two CTF formats represents one test sample with 55 NTF values corresponding to each KU. The test indicated that the distribution difference of KUs between \jeopardy\ and \attackdefense\ format is significant ($U = 1184,\, p = 0.02$). However, we could not use this test to examine the distribution of KAs between the formats. This is because the Mann-Whitney U test requires at least 20 observations in each sample~\cite{PythonMannWhitneyU}, but we have only eight observations (KA values).

\subsection{RQ3: Distribution of Knowledge Areas and Units from 2012 to 2020}
\label{subsec:results-rq3}

This section answers the third research question. We look at the variance in the distribution of KAs and KUs in the CTF writeups divided by year.

\subsubsection{Distribution of Knowledge Areas per Year}
\label{subsubsec:rq3-KA}

\Cref{result:rq3-KAs} shows that the distribution of KAs varies only slightly over the years. However, the chart highlights some deviations.
\textit{Data security} had a lower occurrence in 2012 and 2013.
Similarly, \textit{Software security} had the smallest presence in 2013 and 2014.
However, \textit{Component security} was the most popular in 2013.
\textit{Connection security} is steadily between 17--22\% since 2015.
\textit{System security} peaked in 2014 and stayed at 7--15\% in other years,
with \textit{Human} and \textit{Organizational security} repeating a similar trend.
As for the \textit{Societal security}, it reached its highest percentage, 4.5\%, in 2014 too.

\definecolor{col11}{HTML}{22ABB0}
\definecolor{col12}{HTML}{FAF7EC}
\definecolor{col13}{HTML}{7980E6}
\definecolor{col14}{HTML}{C1FEBB}
\definecolor{col15}{HTML}{AC08CD}
\definecolor{col16}{HTML}{92BFEE}
\definecolor{col17}{HTML}{FFA500}
\definecolor{col18}{HTML}{FF625D}

\begin{figure}[!ht]
\centering
\begin{tikzpicture}
  \begin{axis}[
    ybar stacked,
    ymax=100,
    ytick={20,40,60,80,100},
    ylabel={Percentage},
    yticklabel={\pgfmathprintnumber\tick\%},
    symbolic x coords={2012,2013,2014,2015,2016,2017,2018,2019,2020},
    xtick=data,
    legend style={at={(0.48,-0.20)}, anchor=north,inner sep=1pt, style={column sep=0.1cm, row sep=0.1cm, draw=none}, anchor=north,legend columns=-1},
    legend columns = 4,
    legend cell align={left},
    x=1.2cm,
	bar width=15pt,
    enlarge y limits=0.1,
    ]

  \addplot+[ybar, black, fill=col13] plot coordinates {(2012,18.18) (2013,16.8) (2014,27.96) (2015,29.78) (2016,30.17) (2017,27.18) (2018,26.3) (2019,27.21) (2020,28.1)};
  \addplot+[ybar, black, fill=col14] plot coordinates {(2012,6.43) (2013,3.02) (2014,2.79) (2015,7.57) (2016,8.95) (2017,8.66) (2018,10.85) (2019,11.06) (2020,10.15)};
  \addplot+[ybar, black, fill=col16] plot coordinates {(2012,15.78) (2013,24.67) (2014,4.01) (2015,12.18) (2016,9.02) (2017,11.31) (2018,7.04) (2019,8.87) (2020,8.84)};
  \addplot+[ybar, black, fill=yellow] plot coordinates {(2012,25.04) (2013,28.39) (2014,14.13) (2015,17.17) (2016,18.48) (2017,19.24) (2018,20.54) (2019,19.47) (2020,20.26)};
  
  \addplot+[ybar, black, fill=col18] plot coordinates {(2012,14.6) (2013,7.69) (2014,16.52) (2015,12.98) (2016,13.86) (2017,11.55) (2018,14.13) (2019,12.18) (2020,11.74)};
  \addplot+[ybar, black, fill=col11] plot coordinates {(2012,10.5) (2013,10.21) (2014,14.05) (2015,9.31) (2016,7.67) (2017,8.08) (2018,8.69) (2019,8.8) (2020,7.2)};
  \addplot+[ybar, black, fill=col12] plot coordinates {(2012,7.42) (2013,7.57) (2014,16.01) (2015,7.75) (2016,8.51) (2017,10.55) (2018,9.79) (2019,9.41) (2020,10.97)};
  \addplot+[ybar, black, fill=col17] plot coordinates {(2012,2.06) (2013,1.65) (2014,4.53) (2015,3.26) (2016,3.33) (2017,3.44) (2018,2.67) (2019,2.99) (2020,2.75)};
  
\legend{\strut Data sec.,\strut Software sec.,\strut Component sec.,\strut Connection sec., \strut System sec., \strut Human sec.,\strut Organizational sec.,\strut Societal sec.}
\end{axis}
\end{tikzpicture}
\caption{Distribution of KAs in years 2012--2020.}
\label{result:rq3-KAs}
\end{figure}
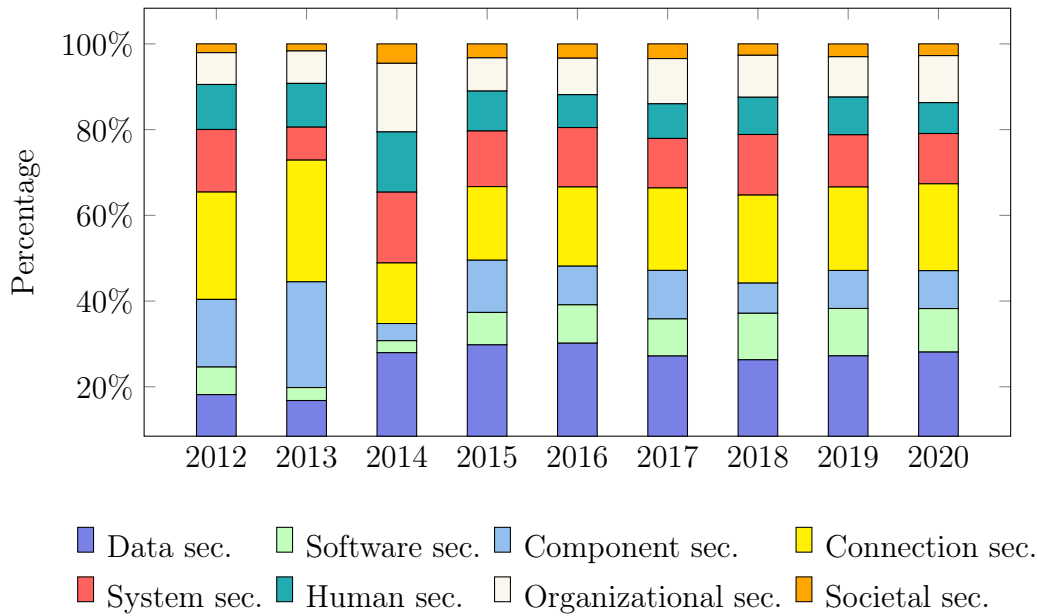


\subsubsection{Distribution of Knowledge Units per Year}
\label{subsubsec:rq3-KU}

Top ranking KUs do not vary much. Between the years 2012--2020, 15 different KUs appeared at the top ten spots, and 12 of them appeared more than once. The steady ones appearing each year are \textit{Cryptography}, \textit{Component design}, and \textit{Network implementations}. Those that appeared every year except one are \textit{System control}, \textit{Digital forensics}, \textit{Distributed systems architecture}, \textit{Network services}, and \textit{Systems administration}. Finally, \textit{Implementation} appeared all years except 2013 and 2014.

We tested the null hypothesis that the medians of KU values throughout 2012–2020 are equal to see if there are statistically significant differences. We used Kruskal-Wallis test~\cite{PythonKruskal} to examine the differences in the distribution of KUs, running it on nine samples (years 2012--2020) with 55 observations (KUs). The test concluded that the difference was significant ($\chi^2 = 25.88,\, p = 0.0011$). When running the test for KA distributions, the test stated that the distribution differences are not significant ($\chi^2 = 0.26,\, p = 0.9999$). Based on \Cref{result:rq3-KAs}, this was expected since the differences appear small.

\subsection{Limitations of This Study}
\label{subsec:limitations}

The main limitation explained in \Cref{subsec:downloading-writeups} is that we downloaded only \numdownloadedwriteups\ out of \numtotalwriteups\ available writeups. Moreover, out of the \numdownloadedwriteups\ analyzed writeups, only 8,688 (54\%) contained at least one keyword. This effectively reduced our dataset, although it remained substantially large. The median length of a writeup was 309 characters, and the average length was 3,979 characters.

Another limitation is the possibility of false positives in the keyword matches. However, we randomly selected 80 writeups (about 1\% of the 8,688) and manually searched for the cybersecurity keywords in their content. Then, we compared these findings with the results of the automated keyword analysis. The results were the same: the desired keywords were matched, and no false positives occurred. If more writeups were validated like this, we would gain greater assurance that no false positives are in the sample, but the process would be time-consuming and error-prone.

Finally, some may find it problematic that the writeups rarely include the precise challenge assignments, which would allow us to double-check the relevance of the writeup to the challenge. However, the writeups were written by experts and enthusiasts who solved the CTF challenge, and so we consider them reliable. In the future, writeup databases could include a separate record of the assignment.

\section{Lessons Learned and Future Work}
\label{sec:lessons}

We now share the educational implications of the results, their comparison with previous work, and practical insights stemming from this research. Finally, we propose ideas for future work.

\subsection{Implications for Cybersecurity Professionals and Educators}
\label{subsec:edu}
Cybersecurity topics are represented unevenly in CTF challenges. This is understandable since some topics, such as attacks on encryption or reverse engineering of code, have been a staple in CTF challenges for years. These technical challenges are popular among CTF creators and participants, who are usually cybersecurity enthusiasts from a private or academic sector. The chosen topics may reflect their opinion on which topics are the most important, are feasible to implement, or are simply fun.

Nevertheless, this opens a new possibility for CTF creators to incorporate human aspects of cybersecurity into the CTF format. Although phishing attacks are a severe threat, as they are the biggest malware infection vector~\cite{enisa-threat-landscape}, our results indicate that the CTF format does not address this topic. Therefore, preparing challenges that teach these aspects can be a valuable and engaging experience, even for those without a deep technical background. This can open up the CTF format for beginners and other users who are not full-time computing experts.

Cybersecurity teachers can use CTF challenges as a suitable hands-on complement to their traditional classes, as was previously tried in~\cite{vykopal2020}. Their regular classes may then focus on areas not covered by CTFs. Moreover, CTFs are excellent for all forms of online learning, including distance and blended learning, due to their remote accessibility. Finally, educators can help their students prepare for CTF challenges. Prominent universities such as Carnegie Mellon or the University of California, Santa Barbara, have their prestigious CTF teams~\cite{ppp, shellphish}.

Starting to participate in CTF is often frustrating for newcomers~\cite{Chung2014Learning, Tobey2014Engaging} since the challenges tend to be difficult. Knowing which topics are the most frequent may help beginners to prepare and avoid disappointment. It may even provide the basis for a CTF training program that would generate recommendations for personalized training paths.

\subsection{Comparison with the Results of Related Publications}
\label{subsec:comparison}

We discovered that \textit{Cryptography} is the most prominent KU overall. Similarly, Burns et al.~\cite{Burns2017Analysis} also identified \textit{crypto} as the top-ranking category of CTF challenges, even though they categorized the writeups differently, not using the cybersecurity curricula.

Cabaj et al.~\cite{Cabaj2018Cybersecurity} found that most cybersecurity master's degree programs include \textit{Data security} (e.g., KU Cryptography), \textit{Software security} (e.g., KU Programming robustly),\textit{ Connection security} (e.g., KU Network defense), and \textit{System security} (e.g., KU Penetration testing). This corresponds to the most prominent topics we found in CTF, and also to those most researched in cybersecurity education papers~\cite{Svabensky2020}. Among the non-technical aspects, \textit{Organizational} and \textit{Societal security} prevail.

The survey of 104 European master's programs in cybersecurity~\cite{CS4EUreport} revealed that Data and Connection security dominate. KUs covered in mandatory courses by more than half of the surveyed institutions were \textit{Cryptography}, \textit{Secure communication protocols}, and \textit{Network defense}. When including also optional courses or subtopics of other courses, these KUs were prominent as well: \textit{Data integrity and authentication}, \textit{Access control}, and \textit{System access}. The least covered KA was \textit{Component security}. Again, these results are similar to ours.

For cyber warfare operations, Carroll~\cite{Carroll2018Offensive} prioritizes networking, fundamental security principles, telecommunications, network defense, and management of vulnerabilities and risks. Most of these topics belong to the \textit{Connection security} KA, followed by \textit{Data security} and \textit{Organizational security}. According to Jones et al.~\cite{Jones2018Core}, cybersecurity professionals also prioritize \textit{Connection} and \textit{System security}.

Overall, \textit{Data} and \textit{Connection security} dominated both in related work and CTFs. These areas include essential technical foundations of cybersecurity, which supports the fact that CTF aligns well with formal study programs. However, educators should not forget about the importance of non-technical aspects, such as human security, privacy, ethics, and law.

\subsection{Legal Aspects of Research That Involves Third Party Data}
\label{subsec:legal}

CTFtime website states that all writeups are copyrighted by their authors. Still, it is allowed to analyze the data for research purposes and present aggregate results as in this paper. In the USA, this is granted by the Copyright Law~\cite{US-copyright-law}. In the European Union, the same exception for research holds~\cite{EU-copyright-law}.

The best practice in research is to publish the analyzed data and software as supplementary materials with the paper. However, this would require obtaining permission from the author of each writeup, which is practically impossible. Without this permission, we would be re-publishing the content, which the copyright does not allow. Therefore, we publish the writeup folder structure, but the writeup files include only the link to the original writeup. Similarly, we cannot publish the Python script used to download the writeups because it would create a local copy of the data unauthorized by the authors. However, it is a simple web scraper that can be replicated based on \Cref{sec:methods}. We publish only the analysis script, which is more specific for this work.

As a guideline for other researchers, we recommend carefully reviewing the conditions under which it is permissible to publish third party materials. In these cases, the support of replicable scientific research and the right to open access to information clashes with the protection of intellectual property. However, if the writeup authors and portals such as CTFtime used a Creative Commons license instead of the traditional copyright, this would simplify the future (re)use of their work.

\subsection{Future Work}
\label{subsec:future-work}

Future researchers can address the limitations of this study, such as downloading the remaining writeups, and thus improve the accuracy of our results. Probabilistic methods can be employed to match the keywords in writeups stochastically. Another possibility is to apply machine learning to the dataset. In~\cite{Rouly2015}, the authors used clustering to \enquote{identify groupings of similar documents according to their term frequency vector Euclidean distance}. The same method can be applied to writeups. If it reveals clusters of writeups on a single topic, it can support the validity of our results. Finally, classification algorithms can categorize the writeups and compare them with the CSEC2017 Knowledge Areas. This fine-grained classification would allow mapping the CTF topics onto specific learning outcomes.

\section{Conclusion}
\label{sec:conclusion}

This work is a pioneering attempt to connect two different aspects of cybersecurity education: (i) popular hands-on challenges prepared by security experts and (ii) formal study programs facilitated by professional educators. If the goal is to exercise cybersecurity skills, CTF challenges suitably complement traditional formats of education delivered by schools and universities. They allow hundreds of students to practice a wide variety of cybersecurity skills online in a hands-on and engaging way.

We analyzed the cybersecurity topics in almost 16,000 written solutions (writeups) of CTF challenges held in the recent decade. The goal of the analysis was to determine how the topics defined in the current Cybersecurity Curricular Guidelines (CSEC2017) are represented in CTF challenges. The analysis showed that topics such as cryptography and network security dominate. Interestingly, the same topics are prevalent in the current study programs and reflect the contemporary literature.

Although CTF challenges are excellent for practicing technical skills, they do not address topics such as phishing and general cybersecurity awareness. However, these topics are of utmost importance for mitigating the current advanced cyber threats. CSEC2017 defines cybersecurity as \enquote{a computing-based discipline involving technology, people, information, and processes to enable assured operations}~\cite{curricula2017}, but the interdisciplinary \enquote{people} aspect is currently missing in CTF. This opens up a new opportunity to design CTFs that would reach a broader, non-technical audience and perhaps attract more people into cybersecurity.

Our paper provides numerous contributions for cybersecurity professionals, teachers, educational managers, CTF participants, and CTF designers. First, we gathered insights into cybersecurity topics practiced via CTF. Next, we discussed the implications of these results and connected them to the state-of-the-art curricular development in cybersecurity. Finally, we provided recommendations for other researchers, along with the directions for future work to motivate further research.

Supplementary materials for the paper are freely published on Zenodo~\cite{dataset}. The archive includes mainly the URLs of the analyzed writeups and the analysis script. This will support other researchers in replicating our work and building upon the analysis of CTF writeups.

\section*{Acknowledgements}
This research was supported by the ERDF project CyberSecurity, CyberCrime and Critical Information Infrastructures Center of Excellence (No. CZ.02.1.01/0.0/0.0/16\_019/0000822).


\bibliographystyle{model1-num-names}
\bibliography{references.bib}

\end{document}